\documentclass{emulateapj}

\usepackage{apjfonts}

\newcommand{\src}{G1.9+0.3}

\catcode`\@=11
\newcommand{\gapprox}{\mathrel{\mathpalette\@versim>}}
\newcommand{\lapprox}{\mathrel{\mathpalette\@versim<}}
\newcommand{\propapprox}{\mathrel{\mathpalette\@versim\propto}}
\newcommand{\@versim}[2]
  {\lower3.1truept\vbox{\baselineskip0pt\lineskip0.5truept
\ialign{$\m@th#1\hfil##\hfil$\crcr#2\crcr\sim\crcr}}}
\catcode`\@=12

\shorttitle{YOUNGEST GALACTIC SNR G1.9+0.3}

\begin{document}

\title{The Youngest Galactic Supernova Remnant:  G1.9+0.3}


\author{Stephen P. Reynolds,\altaffilmark{1}
Kazimierz J. Borkowski,\altaffilmark{1}
David A. Green,\altaffilmark{2}
Una Hwang,\altaffilmark{3}
Ilana Harrus,\altaffilmark{3}
\& Robert Petre \altaffilmark{3}}

\altaffiltext{1}{Department of Physics, North Carolina State University,
  Raleigh NC 27695-8202; stephen\_reynolds@ncsu.edu} 
\altaffiltext{2} {Cavendish Laboratory; 19 J.J. Thomson Ave., 
Cambridge CB3 0HE, UK}
\altaffiltext{3}{NASA/GSFC, Code 660, Greenbelt, MD 20771}

\begin{abstract}

Our 50 ks Chandra observation of the small radio supernova remnant
(SNR) G1.9+0.3 shows a complete shell structure with strong bilateral
symmetry, about $100''$ in diameter.  The radio morphology is also
shell-like, but only about $84''$ in diameter, based on observations
made in 1985.  We attribute the size difference to expansion between
1985 and our Chandra observations of 2007.  Expansion is confirmed in
comparing radio images from 1985 and 2008.  We deduce that \src\ is of
order 100 years old -- the youngest supernova remnant in the Galaxy.
Based on a very high absorbing column density of $5.5 \times 10^{22}$
cm$^{-2}$, we place \src\ near the Galactic Center, at a distance of
about 8.5 kpc, where the mean remnant radius would be about 2 pc,
and the required expansion speed about $14,000$ km s$^{-1}$.  The
X-ray spectrum is featureless and well-described by the exponentially
cut off synchrotron model {\tt srcut}.  With the radio flux at 1 GHz
fixed at 0.9 Jy, we find a spectral index of $0.65$ and a rolloff
frequency of $1.4 \times 10^{18}$ Hz.  The implied characteristic
rolloff electron energy of about $94 (B/10 \ \mu{\rm G})^{-1/2}$ TeV
is the highest ever reported for a shell supernova remnant.  It can
easily be reached by standard diffusive shock acceleration, given the
very high shock velocities; it can be well described by either
age-limited or synchrotron-loss-limited acceleration.  Not only is
\src\ the youngest known Galactic remnant, it is also only the fourth
Galactic X-ray synchrotron-dominated shell supernova remnant.

\end{abstract}

\keywords{
supernova remnants, X-rays : general ---
supernova remnants: individual (\objectname{G1.9+0.3}) ---
X-rays: ISM
}

\section{Introduction}
\label{intro}

Estimates for the supernova rate in the Milky Way Galaxy give about 3
per century \citep[e.g.,][]{vandenbergh91}, so that we expect 60
supernova remnants (SNRs) younger than 2000 yr, while fewer than 10
are confirmed.  This well-known deficit motivated our program of {\sl
Chandra} observations of compact radio remnants, in search of evidence
of youth.  Here we report a confirmation of a young SNR, but one of
remarkable properties: comparison of radio and X-ray images separated
by 22 years demonstrates significant expansion, allowing the inference
of an age of order 100 years.  In addition, its X-ray emission appears
to be purely synchrotron radiation, making G1.9+0.3 the fourth clear-cut
X-ray-synchrotron-dominated Galactic SNR, after SN 1006 \citep{koyama95},
G347.3--0.5 \citep{koyama97}, and G266.2--1.2 \citep{slane01}.

\begin{figure}
\epsscale{1.10}
\plotone{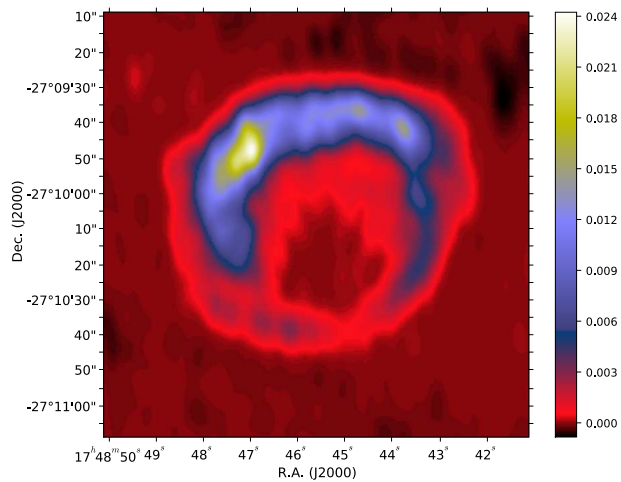}
\caption{1985 radio image of \src\ (Green 2004): VLA at 1.5 GHz.
Resolution $8.4'' \times 3.6'',$ position angle $3.5^\circ$ E of N.
Intensities in Jy/beam.
\label{radioim}}
\end{figure}

G1.9+0.3 was identified by \cite{green84} as a potential young SNR,
from a radio image showing a $1\farcm2$-diameter shell.  This angular
size made \src\ the smallest Galactic SNR, with a mean radio surface
brightness about half that of Tycho's SNR and about 20 times that of
SN 1006, suggesting an age of order $10^3$ yr.  Green \& Gull (1984)
point out the general resemblance of \src\ to Kepler's SNR, and
mention that it would be below 10 pc in diameter at any distance in
the Galaxy.  VLA\footnote{The Very Large Array is operated by the
National Radio Astronomy Observatory, a facility of the National
Science Foundation operated under cooperative agreement by Associated
Universities, Inc.}  observations from 1985 (Green 2004, reproduced in
Fig.~\ref{radioim}) show a strong asymmetry in the shell at 21 cm,
perhaps indicative of an external density gradient.  Radio flux
measurements with different instruments and frequencies are
inconsistent; from VLA measurements, \cite{nord04} report a radio
spectral index $\alpha \sim 0.7$ ($S_\nu \propto \nu^{-\alpha}$)
between 74 and 327 MHz (consistent with flux estimates by Gray 1994)
and giving a 1 GHz flux of 0.9 Jy.  This steep a spectrum is typical
of historical SNRs but rare otherwise.  No counterparts are
obvious in optical or infrared images.  The remnant was detected with
{\it ASCA} \citep{sakano02}, but contamination from nearby bright
sources precluded detailed analysis.

\begin{figure}
\epsscale{1.10}
\plotone{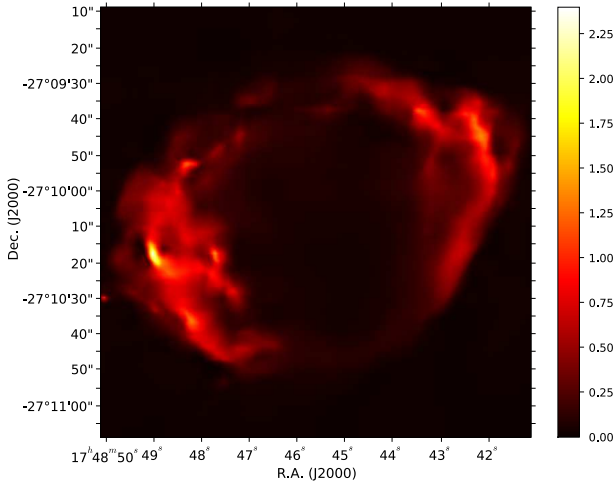}
\caption{{\sl Chandra} image of \src, platelet smoothed,
covering the same area as Figure 1.  
Colors are counts/ACIS pixel ($0.49''$)between 1.5 and 6 keV.
\label{xim}}
\end{figure}

\section{X-ray Images and Spectra}

We observed \src\ with {\sl Chandra} using the ACIS-S CCD camera (S3
chip) on 2007 February 10 for 24 ks, and March 3 for 26 ks.  We
checked aspect correction and created new level--1 event files
appropriate for VFAINT mode. No flares occurred during the
observation.  CTI correction was applied and calibration was performed
using CALDB version 3.4.0.  Finally, the datasets were merged and
weighted response files created. We extracted spectra using the {\tt
specextract} script, using a background region on the S3 chip.  About
8,000 counts were obtained from the remnant.

Figure~\ref{xim} shows the image over the full energy range,
platelet-smoothed (method described in Willett 2007).  The bright ring
is nearly circular, though fainter extensions (``ears'') protrude
symmetrically on the E and W sides.  The azimuthal brightness
variations are quite different in radio and X-rays; the former has a
single maximum to the north, while the latter shows remarkable
bilateral symmetry reminiscent of SN 1006. However, the difference in
radio and X-ray morphologies is quite unlike SN 1006.  The angular
extent is considerably larger than suggested by the radio image: about
$120'' \times 90''$.  The bright ring is slightly smaller E-W, about
$100'' \times 90''$.  No central source is apparent.

\begin{figure}
\vspace{-0.1in}
\epsscale{0.7}
\plotone{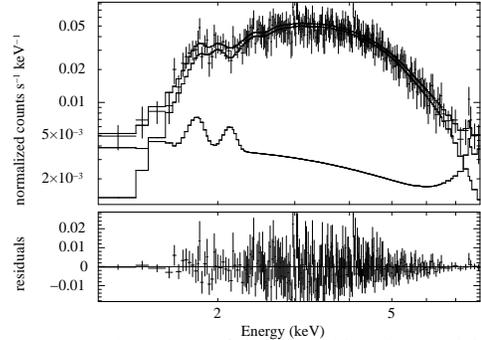}
\vspace{-0.1in}
\caption{Integrated spectrum of \src, binned to a minimum of 25
counts per channel for display purposes only.  Model: {\tt srcut} (see text).
 \label{totspec}}
\end{figure}

Instead of subtracting background (rendering the statistics
non-Poisson), we modeled it with a combination of Gaussians and
power-laws, exponentially cut off on both ends.  The spectrum is shown
in Figure~\ref{totspec}, along with the background model.  No lines at
all are evident.  Very high absorption is also apparent.  We fit the
data with the {\tt srcut} model for synchrotron emission from a
power-law distribution of electrons with a high-energy exponential
cutoff (Reynolds \& Keohane 1999).  The model has four parameters:
absorbing column density $N_H$, 1-GHz radio flux, power-law spectral
index $\alpha$, and rolloff frequency $\nu_{\rm roll}$; we fixed the
radio flux to 0.9 Jy (though brightening may have occurred; Green et
al.~2008).  Parameter estimation was done with Markov Chain Monte
Carlo simulations, as implemented in the XSPEC package
\citep{arnaud96}.  Constant Bayesian priors were assumed for the three
undetermined parameters. We found $N_H = (5.5 \pm 0.3) \times 10^{22}$
cm$^{-2}$, $\alpha = 0.65 \pm 0.02$ and $\nu_{\rm roll} =
1.4_{-0.7}^{+2.1} \times 10^{18}$ Hz ($h\nu_{\rm roll} =
5.8_{-2.9}^{+8.7}$ keV) -- one of the highest values ever reported for
a SNR.  (Errors are 90\% confidence intervals.) Now the high column
means that dust scattering will be important, preferentially for
lower-energy X-rays, hardening the spectrum.  Including scattering
will reduce the inferred column density and rolloff frequency
somewhat.  Detailed modeling will be required for quantitative
assessment.  In any case, evidently we have discovered the fourth
example of a Galactic synchrotron-X-ray dominated SNR.

\begin{figure}
\epsscale{0.8}
\plotone{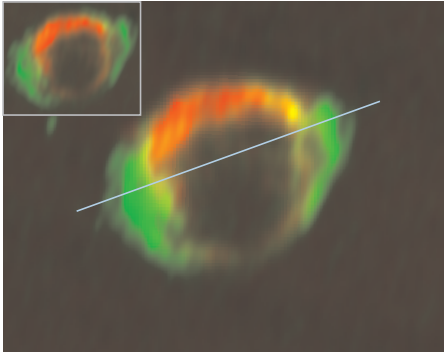}
\caption{Comparison of radio images (red) and X-ray image smoothed
to $10'' \times 4''$. Inset:  Original 1985
radio image.  Main figure:  Radio image expanded by 16\%.
Lines: orientation of profile in Fig.~5.}
\label{x-rcomp}
\end{figure}

The striking size difference led us to compare the radio and X-ray
images in more detail.  The bright X-ray ring shows indentations and
bright features that correlate well with the outer extent of the radio
image -- though at considerably larger radius, as can be seen in
Figure~\ref{x-rcomp}.  We smoothed the raw X-ray image to the
resolution of the radio image, and examined an E-W profile (Figure~5)
chosen to pass through prominent radio and X-ray features.  Finite
resolution should not affect the radio/X-ray size ratios.  The
distance between X-ray peaks, about $95''$, should be a lower limit to
the X-ray diameter; if the X-rays abruptly turned on at the shock,
those locations would appear as the outer inflection points, separated
by about $100''$, a clear upper limit.  The outer radio edge was
estimated from inflection points in the radio profile, not from the
peaks, as we expect an intrinsically much thinner X-ray than radio
ring.  The radio diameter judged in this way was $84''$.  These values
give fractional expansions of between 13\% and 19\% of the 1985 size.
We adopt an estimate of $(16 \pm 3)$\% for the expansion in the 22
years since 1985, giving an age estimate of about $140 \pm 30$ yr, or
less if deceleration has occurred as is likely.  We shall argue below
that \src\ is near the center of the Galaxy; at 8.5 kpc, the shock
speed is 14,000 km s$^{-1}$.  While such speeds are seen in optical
spectra of supernovae, they are unprecedented in supernova remnants.

We recently obtained VLA Exploratory Time to examine \src\ at the
present epoch (Green et al.~2008).  The VLA is currently in C
configuration, so we observed at 6 cm to achieve comparable
resolutions ($10'' \times 4'')$.  Averaged over all azimuths, the
shell profile peaks at a radius of about $30''$ in 1985 and $34.5''$
in 2008.  There are also some morphological differences in the images;
however, a careful analysis of the averaged profiles gives an
expansion of $(15 \pm 2)$\% (Green et al.~2008), consistent with our
determination.

\begin{figure}
\epsscale{0.8}
\plotone{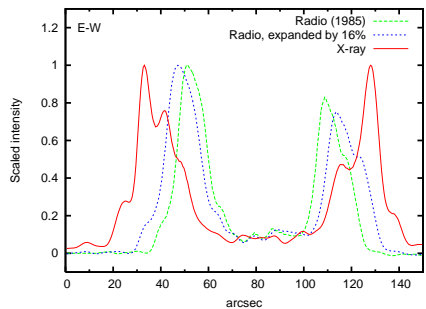}
\caption{Profiles of radio and X-ray images.  Outer (red) line: X-ray.
Dashed green line: Original radio image.  Short-dashed blue line:
Radio expanded by 16\% about the center of the ring, to bring outer
inflection points to X-ray maxima.  X-ray extensions (``ears'') are
apparent at about 30\% of maximum.}
\vspace{-0.1in}
\label{profile}
\end{figure}

\section{Discussion}

The extremely high absorption toward \src\ makes it unlikely that it
is much closer than the Galactic Center (GC) region, at about 8.5 kpc.
In fact, the Galactic extinction model of \citet{marshall06}, based on
near infrared 2MASS extinction colors, gives an infrared absorption of
$A_{K_s} = 1.8^m$ at the GC distance and $l=1.87^\circ,
b=0.32^\circ$. Assuming the same amount of heavy-element depletion
onto dust grains as in the dust model of \citet{wg01}, about half of
the absorption can be accounted for by material located in front of
the GC.  However, the very high expansion we infer argues against much
larger distances than to the Galactic Center, suggesting that the
additional absorption is local to the source.  We assume a nominal
distance of 8.5 kpc, at which $1'' = 0.041$ pc.  Then the mean radius
of the bright X-ray ring is about 2 pc, with the E and W ``ears'' at
about 2.2 pc.  The very high expansion velocities would be mitigated
at a substantially smaller distance, but the extreme absorption and
absence of any IR counterpart (for instance, from a more local dense
cloud of some kind) argue strongly against this.

Our inferred age of about 100 yr makes \src\ the youngest known
remnant in the Galaxy.  The very high absorption would have rendered
the supernova undetectably faint.  The average optical absorption
$A_V$ foreground to the GC exceeds $15^m$, and the dust local to the
GC region contributes a comparable amount, resulting in $A_V>30^m$.
Even a bright SN Ia with $M_V \sim -19^m$ would be fainter than $m_V
\sim 25^m$.

Three arguments slightly favor a Type Ia origin for \src.  First, the
bilateral symmetry of the X-ray synchrotron emission suggests
interaction with a roughly uniform magnetic field, with electron
acceleration dependent on the obliquity angle between shock normal and
{\bf B} (e.g., Fulbright \& Reynolds 1990).  A stellar-wind magnetic
field, as expected for a core-collapse (CC) progenitor, should be a
tightly wound Archimedean spiral, i.e., almost perpendicular to the
shock normal at all azimuths, removing that possible explanation for
the nonthermal morphology.  However, Tycho, a prototype Ia remnant,
does not show bilateral symmetry, nor does the radio emission in
\src.  Second, the very high expansion speeds are slightly more
probable for a Ia supernova.  Third, there is no indication of a
pulsar-wind nebula at the center (but a compact object with a soft
spectrum like that in Cas A could not be seen with this high
absorption).  The question is still open; the position of \src\ near
the GC is consistent with either type, as both the bulge and disk
populations are substantial there.

The high velocities suggest that little deceleration has occurred.
The mass of gas in a sphere of radius 2.2 pc is only $1.5n_0$
$M_\odot$ for a mean number density $n_0$ and cosmic abundances.  A
model Type Ia explosion with an exponential ejecta profile
\citep{dwarkadas98} achieved the current size and shock velocity for
standard explosion parameters, an age of 100 yr, and $n_0 \sim
4 \times 10^{-2}$ cm$^{-3}$, characteristic of the ISM hot phase.  For this
density, the extremely low X-ray emission measure, combined with the
ionization timescale $\tau = n_0 t$ of only $\sim 10^8$
cm$^{-3}$ s, means that no significant X-ray line emission would be
expected from shocked ISM.  The age is so small that very little of
the ejecta have been shocked, so a contribution to the spectrum from
shocked ejecta should be small as well.

Densities might be larger than average in the northern radio-bright
limb. We partitioned the shell into two sections, the northern limb
and the remainder, and made joint fits with a nonthermal {\tt srcut}
model in both limb sections, adding a {\tt vpshock} model with cosmic
abundances in the northern limb alone. Thermal emission with emission
measure as high as $1.6$ $M_\odot$ cm$^{-3}$ for gas with temperature
4 keV and shock age $2 \times 10^9$ cm$^{-3}$ s can be accommodated by
the current data; the very low ionization age suppresses most X-ray
lines.  (However, the absence of Fe K$\alpha$ emission confirms that
the continuum we see is nonthermal.)

The rolloff photon energy is related to the e-folding energy of the
exponential cutoff in the electron spectrum, $E_m$, by $E_m = 39
(h\nu_{\rm roll})^{1/2}B_{10}^{-1/2}$ TeV \citep[][with a numerical
error corrected]{reynolds99}, where $h \nu_{\rm roll}$ is in keV, and
$B_{10} \equiv B/10 \ \mu$G is the post-shock magnetic field.  From
the fitted value $h \nu_{\rm roll} = 5.9$ keV, we infer $E_m = 94$ TeV
-- the highest value ever reported for an integrated spectrum.
However, with our very high shock velocities, the predictions of
standard shock acceleration theory can easily accommodate such
energies.  If radiative losses cut off the electron spectrum, $E_m
\propto B^{-1/2}$, so $h \nu_{\rm roll} = 25.7 E_m^2 B$ keV is
independent of $B$.  Standard results (e.g., Reynolds 1998) for a
shock with speed $u_8 \equiv u/1000$ km s$^{-1}$ give $h\nu_{\rm roll}
\sim 0.8 u_8^2$ keV, for Bohm diffusion and a perpendicular shock, and
a compression ratio of 4 (the estimate depends only weakly on these
assumptions).  We estimate $u_8 = 14$, so we can easily reach the
required value, even with considerably less efficient acceleration.
An estimate of the limitation due to finite age can be made assuming
no deceleration; the expression for acceleration time to energy $E$
given in Reynolds (1998) gives $\tau(94\ {\rm TeV}) = 300 B_{10}^{-1}
u_8^{-2}$ yr, so that the shock could produce our observed $E_m$ in
only a few years.  Likely higher values of $E_m$ at the bright limbs
would then not pose any problem.

The contrasting X-ray and radio azimuthal brightness profiles imply
substantial spectral differences between the northern and southern
hemispheres.  Preliminary spectral fits do suggest differences, but
unexpectedly mainly in $\alpha$ rather than $\nu_{\rm roll}$.
Efficient shock acceleration is expected to produce concave-up
curvature in the particle spectrum (hardening to higher energies), so
that $\alpha$ in {\tt srcut} is a mean radio--to--X-ray spectral
index.  If deeper X-ray observations confirm this form of spectral
difference, the result may have significant implications for the
understanding of shock acceleration.  Other radio--X-ray differences,
such as the faint ``ears'' of X-ray emission beyond the bright ring on
the E and W sides, could in principle be due to pre-shock diffusion,
or simply reflect lower sensitivity to faint radio emission.
Estimates of the diffusion length, assuming a mean free path of one
gyroradius of a 94 TeV electron in a magnetic field of a few
$\mu$Gauss, are roughly consistent with the observed length scale of
$5'' - 10''$, but more extensive radio and X-ray observations are
needed to test this explanation.

A supernova this recent should produce a significant signal in the
1.156 MeV gamma-ray decay line as radioactive $^{44}$Ti decays first
to $^{44}$Sc and then to $^{44}$Ca, as has been reported for Cas A
\citep{iyudin94}.  Estimates of the yield of $^{44}$Ti in SNe Ia range
from $8 \times 10^{-6}\ M_\odot$ to $5 \times 10^{-5} \ M_\odot$
\citep{iwamoto99}, while CC predictions range between $ 2 \times
10^{-5}$ and $2 \times 10^{-4} \ M_\odot$ \citep{thielemann96}.  For a
half-life $t_{1/2} = 60.3 \pm 1.3$ yr \citep{goerres98}, we predict a
flux of $2.3 \times 10^{-6} (M({\rm Ti})/10^{-5} \ M_\odot)$ ph
cm$^{-2}$ s$^{-1}$ at an age of 140 yr. For comparison, the flux
reported for Cas A is $3.4 \times 10^{-5}$ ph cm$^{-2}$ s$^{-1}$
\citep{iyudin94}.  \citet{renaud06} examined INTEGRAL data from the GC
region, acquired within the first 2 years of its operations, while
searching for $^{44}$Ti-decay emission from another claimed young
supernova remnant, G0.570--0.018.  They showed a map of the Galactic
plane near the GC with no evidence for emission from \src.  They did
not detect their target; the limit should apply as well to \src.  For
an age of 140 yr, their $3\sigma$ upper limit on the yield of
$^{44}$Ti is $4 \times 10^{-5} \ M_\odot$; at 160 yr, it has increased
to $6 \times 10^{-5} \ M_\odot$.  We believe a deeper INTEGRAL
observation will provide an important test of the youth of \src\, and
may even help discriminate among progenitor masses (for CC) or
explosion mechanisms (for Ia).

We can make crude predictions of continuum gamma-ray fluxes to be
expected from \src, using results quoted in Reynolds (2008).  The
integral photon flux above 100 MeV due to the decay of $\pi^0$ mesons
created by the collisions of cosmic-ray ions accelerated in \src\ with
ambient thermal gas is about $6 \times 10^{-9} n_0 E_{51} \theta$ ph
cm$^{-2}$ s$^{-1}$, where the explosion energy is $10^{51} E_{51}$
erg, and $\theta \sim 0.1$ is the fraction of explosion energy in
cosmic-ray ions \citep{drury94}. This should very roughly scale as
photon energy $E^{-1}$ or slightly steeper, so that the integrated
flux above 0.2 TeV is about $3 \times 10^{-12} n_0 E_{51} \theta$ ph
cm$^{-2}$ s$^{-1}$, or about 1 mCrab.  This is about $20 n_0$ times
below the H.E.S.S. upper limit in the GC region \citep{aharonian06};
with $\theta = 0.1$ and our density estimate of $n_0 \sim 10^{-2}$
cm$^{-3}$, we do not expect a detectable signal.  The very high $E_m$
inferred from the synchrotron spectrum means that the predicted
inverse-Compton spectrum from upscattering cosmic microwave background
photons (ICCMB) will be depressed by Klein-Nishina effects.  Any TeV
upper limit can be accommodated by simply raising the magnetic field,
as it means that fewer electrons are required to explain the
synchrotron spectrum.  But the predictions for ICCMB TeV flux are so
low that the lower limit on the magnetic field corresponding to the
current H.E.S.S. limit is less than 1 $\mu$G.

\section{Conclusions}

We have detected expansion between 1985 and 2007 that demonstrates
that SNR \src\ has an age of about 100 years.  It also has a
synchrotron-dominated X-ray spectrum with the highest rolloff energy
yet observed, but consistent with this very young age.  Synchrotron
domination requires both high shock speeds, for acceleration to
energies of order 100 TeV and above, and low ambient densities, to
suppress thermal X-ray emission.  Both these conditions appear to be
present for \src.  The implied lack of deceleration is consistent with
the requirement of a lack of thermal emission.

Our results invite many predictions.  Our first, that a new radio
observation should immediately show expansion, has already been
verified.  Another Chandra observation should show it as well.  Time
variations in the nonthermal fluxes, both radio and X-ray, are likely
and may give information about the density profiles of ejecta and/or
ambient medium.  The INTEGRAL satellite will observe \src\ shortly,
testing the prediction for the 1.16 MeV gamma-ray line.  The high
absorbing column makes optical detection hopeless, but the implied
high dust density suggests that a search for infrared light echos,
such as those recently detected from Cas A
\citep{krause05,kim08,dwek08}, should be undertaken.  At a projected
distance of only 300 pc from the GC, \src\ is likely to be near
substantial local dust.  A more extensive X-ray observation to search
for any thermal emission is crucial, as such emission could
corroborate this picture and also give information on a stage of SNR
evolution never before observed.

\acknowledgments

This work was supported by NASA through Chandra General Observer
Program grant GO6-7059X.


\end{document}